# Gender and Neighborhood Penalties in Karachi's Information Technology

# Sana Khalil

**Abstract**

  This paper examines how gender and residential socioeconomic status shape hiring outcomes in Karachi's information technology sector. Employers in Pakistan can openly state preferences regarding gender, residential location, and other characteristics, but the majority in the information technology sector choose not to do so. This creates an opportunity to examine whether discrimination persists when such biases are not explicitly stated.

  An analysis of explicitly gender-targeted job ads shows that men are preferred over women across most occupations, even in traditionally pink-collar roles. Results from a resume audit experiment, submitting 2,032 applications to 508 full-time job openings, show that men receive more callbacks for job interviews than women, even in the absence of explicit gender preferences in job ads.

  The study also indicates a significant premium favoring candidates from high-income areas, who receive 45% more callbacks than applicants from low-income neighborhoods. This advantage remains robust even after controlling for commuting distance.

  Qualitative interviews with human resource officials suggest that employers associate productivity with both gender and neighborhood socioeconomic status. Residential address acts as a proxy for class background and signals education, skills, and perceived "fit" in professional settings. These perceptions may reinforce stereotypes, disadvantaging women and candidates from low-income backgrounds.

## 1. INTRODUCTION

Correspondence studies, in which fictitious applications are sent in response to actual job openings, have revealed instances of hiring discrimination against blacks (Bertrand and Mullainathan, 2004; Nunley et al., 2014; Kline, Rose, and Walters, 2022) and various ethnic minorities (Galarza and Yamada, 2014; Zschirnt and Ruedin, 2016; Weichselbaumer, 2020). Moreover, individuals from other disadvantaged groups, such as transgender individuals (Granberg, Andersson, and Ahmed, 2020; Aksoy et al., 2022), gay men (Tilcsik, 2011; Ahmed et al., 2013), lesbian women (Ahmed et al., 2008; Weichselbaumer, 2003; 2015), disabled individuals (Ameri et al., 2018), ex-offenders (Pager, 2003; Ahmed and Lang, 2017), and older candidates (Ahmed et al., 2012; Riach, 2015; Carlsson and Eriksson, 2019), continue to face discrimination in hiring processes (for a comprehensive overview, refer to Baert, 2018, and Neumark, 2018). Despite this rich literature, relatively little is known about how employers interpret socioeconomic and neighborhood signals during the screening process.

In cities of the Global South, such as Karachi, Pakistan, urban development follows a stark core-periphery model (Rana et al., 2017), with severe inequalities in infrastructure, education, and healthcare. These disparities render residential address a potent proxy for socioeconomic status, productivity, and cultural "fit" in professional settings. For example, employers may view candidates from low-income neighborhoods as riskier or less desirable hires (Tunstall et al., 2014). This study situates its investigation of neighborhood-based hiring penalties within these broader urban inequalities, where location-based cues can shape labor market sorting and reinforce exclusion.

Two theoretical mechanisms explain how such biases can emerge (see Carlsson and Eriksson, 2023, for example). First, employers may rely on easily observable signals—like neighborhood of residence—to statistically discriminate, particularly when certain areas are associated with poverty, crime, or marginalized ethnic groups (Phelps, 1972; Hillier, 2003; Small and Pager, 2020). Second, residential distance from the workplace may raise concerns about lateness, fatigue, or job retention, leading employers to penalize applicants from geographically distant or underserved neighborhoods, even when actual commuting time is accounted for.

Pakistan presents a compelling setting to investigate employers' hiring behavior. With a population exceeding 250 million, it is the second-largest Islamic country in the world,[1] where little is documented about employers' hiring practices (Baert, 2018). Neighborhood signaling effects play a key role in Pakistan's context, where residential address is more than a geographic detail—it serves as a deeply embedded social signal, reflecting and reinforcing broader structures of inequality. As Mallick (2018) explains, Pakistan's post-liberalization urban landscape is increasingly shaped by the proto-hegemonic aspirations of a newly emergent middle class, which expresses its claims to modernity through exclusionary spatial practices such as gated communities and the securitization or displacement of informal settlements. The dominant urban narrative stigmatizes low-income neighborhoods as spaces of disorder, backwardness, and even criminality. This spatial stigma is further evident in Hussain et al.'s (2023) study, which shows that youth from informal settlements in Islamabad are perceived and treated differently based on their residential location, facing social exclusion and criminal labeling. Such stigma is compounded by structural disparities in access to education, infrastructure, and services (Rana et al., 2020). In this context, neighborhood-based hiring penalties not only limit access to employment but can also reinforce residential segregation and constrain intergenerational mobility.

Gender also plays a strong role in the labor market screening processes in Pakistan. In the broader society, patriarchal structures are reinforced through institutionalized gender segregation, and women often face penalties for deviating from rigid gender norms (Khalil, 2024). Factors such as socioeconomic status and residential location significantly influence the diverse experiences and expectations of women in Pakistan (see Shah, 2024, for a detailed discussion of cultural and historical factors). Moreover, due to vague wording in institutional regulations regarding equal employment opportunities[2] and a lack of enforcement, employers can openly state preferences regarding gender, residential location, and other characteristics. This practice is more common in online job postings than in newspapers.[3] While gender preferences in job ads are prevalent in developing countries (Kuhn and Shen, 2013; Chowdhury et al., 2018; Kuhn, Shen, and Zhang, 2020), some aspects of the hiring practices in Pakistan appear distinctive. For example, gender preferences are not only explicitly stated but rigorously enforced. An analysis of job ads posted on Pakistan's largest online job portal, rozee.pk, reveals gender preferences for men, women, and transgender candidates (Khalil, 2021). My correspondence analysis shows that employers strictly adhere to their explicitly stated gender preferences, are more likely to favor men across most of the occupations, and often block applications from candidates of the mismatched gender.

---

[1] https://worldpopulationreview.com/countries/pakistan

[2] Articles 25 and 26 of the Constitution of Islamic Republic of Pakistan 1973 prohibit discrimination based on gender, caste, race, religion, residence or place of birth but make leeway for 'gendered' hiring by suggesting that "specific services can be reserved for members of either sex if such posts/services require duties which cannot be adequately performed by the members of the other sex". Article 27 also adds a similar 'suitability' clause. See https://www.ilo.org/dyn/natlex/docs/ELECTRONIC/33863/76773/F1453295923/PAK33863.pdf

Interestingly, about 80 percent of the employers in the information technology (IT) sector do not specify gender preferences in job ads (Matsuda, Ahmed, and Nomura, 2019)[4]. While this could signal greater gender inclusivity, implicit bias may still be at play (see Bertrand, Chugh, and Mullainathan, 2005). This also creates an opportunity to examine whether patterns of discriminatory behavior persist when such biases are not overtly expressed (in a context where open expressions of preference are otherwise permitted).

This study investigates employers' hiring practices in Karachi's information technology (IT) sector, focusing on how applicants' gender and socioeconomic status—proxied by the income levels of their residential locations—influence the likelihood of receiving a callback for a job interview.[5] Several factors motivate this sectoral focus. First, software programming roles offer high pay, require advanced technical skills, and operate within a highly competitive job market, which tends to reduce the applicant pool. In such settings, employers might prioritize technical competence over social identity factors (Azmat and Petrongolo, 2014). Second, prior research, such as Banerjee et al. (2009), suggests that in skill-intensive fields, employers may be less likely to discriminate based on identity markers. Banerjee et al. (2009) suggest that software programming jobs typically involve minimal face-to-face interaction, diminishing the role of "soft skills" such as interpersonal and language abilities. As a result, there may be less emphasis on social identity characteristics when hiring for these positions. This study aims to test this proposition.

Banerjee et al.'s study in Delhi found that lower-caste applicants faced bias in call centers due to soft skill demands, but not in software jobs where hard skills dominated. In contrast, Thijssen et al. (2021), drawing on correspondence studies from five European countries, find that even when applicants' productivity is signaled through diagnostic information, ethnic discrimination persists (see Neumark and Rich, 2019, for a review of similar studies).

To examine discriminatory hiring practices in Pakistan, this study follows a three-pronged research approach. First, it analyzes job ad data to document explicit gender preferences across occupations. The analysis reveals that gender-based sorting is overtly embedded in hiring practices in Pakistan, systematically excluding women from the majority of occupations. Second, a resume audit is conducted to test whether discrimination persists when employers do not explicitly state biases, in a context where such expressions are otherwise permitted. By randomly varying applicants' gender and residential location—used as a proxy for socioeconomic status—the experiment isolates the effects of these signals on interview callback rates. Following the research design from Bertrand and Mullainathan (2004), resumes of software programmers are compiled from LinkedIn and modified to ensure they are not traceable back to their original owners. Names and residential addresses from both high- and low-income neighborhoods are then randomly assigned to these modified resumes. To further control for any residual differences in resume layout, the resumes are periodically swapped among the fictitious applicants. In the research design, applicants are matched based on their productive characteristics, and their outcomes are compared against the same employer.

The resume audit shows that even when jobs don't indicate any explicit gender preferences, men receive more callbacks for job interviews than women. Moreover, there is a substantial and significant neighborhood penalty against candidates from low-income areas, with a premium for candidates in high-income neighborhoods. Candidates from high-income neighborhoods get 45 percent higher callbacks for job interviews than candidates from low-income neighborhoods. Third, the study conducts qualitative surveys with human resource (HR) officials from IT and Sales and Business Development (SBD) firms to investigate how employers interpret gender and neighborhood information during the screening process. While resume audits offer preliminary evidence of discrimination, they provide limited insight into its underlying causes. These qualitative surveys deepen our understanding of the hiring attitudes of those directly involved in screening and shed light on the beliefs and heuristics that may drive observed patterns, particularly how residential location and gender shape perceptions of professionalism, productivity, and organizational 'fit.' Survey results indicate that HR officials in IT firms are more likely to associate productive traits—such as productivity, punctuality, and professionalism—with neighborhood status, more often than with gender. In contrast, HR officials in SBD firms are more likely to associate negative work traits with a candidate's gender and residential address compared to their counterparts in the IT sector. The survey findings suggest potential avenues for further research, such as examining why HR officials in Sales and Business Development (SBD) firms exhibit more substantial biases.[6] One possible explanation is that in IT jobs, face-to-face interactions are often limited, meaning that biases or discriminatory preferences could arguably be attributed to employer practices. Conversely, positions in SBD firms typically involve greater direct interaction with customers and colleagues. The survey findings align with theoretical predictions from Combes et al. (2016), who propose that customer discrimination represents a distinct form of bias affecting jobs with greater interpersonal contact. Using French Census Data, Combes et al. show that African immigrants face significant underrepresentation in customer-contact positions (10 percentage point gap) and higher unemployment rates (11 percentage point gap), compared to French natives, that cannot be explained by skill differences alone. The unemployment rates increased in areas with higher concentrations of French natives, suggesting that customer preferences may drive discriminatory hiring practices when jobs require direct interaction with clients. My survey findings, in this regard, suggest potential avenues for future research.

This study makes several contributions. It addresses a gap in the literature on developing countries by presenting the first field experiment on hiring discrimination in Pakistan's labor market, examining how gender and socioeconomic signals influence employers' screening process and hiring behavior.[7] Most existing resume audit studies focus on high-income countries with formal labor market structures and legal safeguards against explicit bias. In contrast, this study examines hiring practices in Pakistan's largely unregulated labor market, where employers can openly express discriminatory preferences. This setting provides an opportunity to observe different forms of hiring discrimination and to explore the mechanisms through which it operates.

Second, it triangulates findings across job ad data, audit experiments, and qualitative interviews to examine both explicit and subtle discrimination. The job ad data reveal a strong preference for men across most occupations in Pakistan, particularly in fields such as Engineering and Maintenance/Repair. Even in traditionally pink-collar sectors like Health and Medicine, men are favored (21%) over women (approximately 6%). The resume audit indicates discrimination even when job postings do not explicitly state biases, and qualitative interviews with human resource professionals reveal that employers often associate attributes such as productivity, punctuality, and professionalism with both gender and socioeconomic background. Together, these findings offer a more comprehensive understanding of discriminatory hiring practices in a labor market with limited institutional protections.

Finally, this study contributes to the literature on spatial mismatch and stratification economics by providing evidence that neighborhood affluence influences employers' hiring decisions in a high-skilled labor market, even after controlling for commuting distance. While the literature on residential location's influence on worker economic success is well-established (Sharkey and Faber, 2014; Chyn and Katz, 2021; Chetty et al., 2020), there is limited causal evidence on how socioeconomic signals influence applicants' likelihood to get job interviews. Only a few studies have explored this question, all within developed country contexts (Tunstall et al., 2014; Carlsson, Reshid, and Rooth, 2018;

---

[6] Given the time and resource limitations, along with other specified considerations, this study primarily focused on the IT sector. However, I hypothesize that both gender and residential location may play a more significant role in hiring practices within Sales and Business Development (SBD) firms due to the greater level of interpersonal contact in these positions.

[7] According to Baert's (2018) overview of correspondence experiments conducted over the past decade, Pakistan is among the largest countries by population, where no correspondence experiments on labor market outcomes have been conducted to provide any direct evidence of discrimination.

Phillips, 2020). Carlsson et al. (2018), for example, examined whether living in deprived neighborhoods affects interview callback rates for applicants with Swedish and Middle Eastern names in two major Swedish cities. While they found no neighborhood signaling effect for Swedish names, those with Middle Eastern names from deprived neighborhoods received 42% fewer callbacks compared to counterparts from affluent neighborhoods.

Neighborhood signaling effects are particularly important in Pakistan's context, where political and economic power is concentrated among men and the upper-class elite, and neighborhoods are highly segregated by income. Karachi—the largest metropolitan city in Pakistan, with over 18 million residents—exhibits stark disparities between rich and poor neighborhoods in terms of access to education, infrastructure, and sanitation services (Khalil et al., 2023). In this context, residential location serves as a strong signal of socioeconomic privileges, status, and educational background (see Mahboob, 2017; Shah, 2024, for further context). While much of the literature has explored how neighborhood and residential location shape labor market outcomes via social networks and job referrals (Bailey et al., 2020; Krauth, 2004; Fernandez and Mateo, 2006; Schmutte, 2015; Gee, Jones, and Burke, 2017), research from the demand side remains limited. Some studies suggest that, from the demand perspective, commuting distance is more relevant than neighborhood affluence or socioeconomic status in employers' hiring decisions, as long commutes can signal potential fatigue, low work effort, or higher absenteeism (Phillips, 2020; Diaz and Salas, 2020). However, this study finds that employers in Karachi's IT sector penalize candidates from low-income neighborhoods, even when commute distance is held constant. Two mechanisms likely explain this: (1) neighborhood affluence may signal stronger social networks and embedded resources, prompting employers to favor applicants from high-income neighborhoods to leverage and preserve these networks (Petersen et al., 2000).[8] Second, employers may associate low-income areas with low productivity, leading to statistical discrimination (Diaz and Salas, 2020; Carlsson and Eriksson, 2023).

The findings from this study carry broader policy implications not only for Pakistan but also for other developing countries grappling with intergenerational inequality and limited economic mobility. Even when individuals from disadvantaged backgrounds acquire strong qualifications, neighborhood-based stereotypes may constrain their upward economic mobility. While this study is limited in scope, it highlights the need for further research on how residential segregation shapes hiring practices in developing countries.

The rest of this paper is structured as follows. The next section reviews the related literature, Section 3 lays out the methodology and techniques used in this study, Section 4 discusses the results, and the final section concludes.

## 2. BACKGROUND AND LITERATURE

Two critical issues characterize Pakistan's labor market: persistent gender gaps in labor outcomes and extremely low intergenerational income and occupational mobility. Pakistan has one of the highest gender wage gaps globally, with women earning, on average, 34% less than men (International Labor Organization [ILO], 2018), and gender inequalities in access to schooling remain high (Hasan, Rehman, and Zhang, 2021). While women's literacy and higher education enrollment have improved in recent decades[9], their access to public and private sector jobs remains restricted (Shah, 2024). According to the 2020-21 Pakistan Labor Force Survey, the overall proportion of women in managerial positions is only 5.7%.[10] Among college-educated women, unemployment stands at 43%, compared to just 7% for men (Pakistan Bureau of Statistics, 2018-19). Women's labor force participation rate remains stagnant at 25%—well below the 45% average for middle-income countries.[11]

---

[8] Petersen et al. (2000) provide compelling evidence of how social networks shape hiring outcomes by analyzing all 35,229 job applicants to a midsized high-technology organization over a 10-year period (1985-1994). Their findings reveal stark differences in how gender and race affect the hiring process. For gender, the process appeared meritocratic, with age and education accounting for all sex differences in hiring outcomes. However, for ethnic minorities, the process was only partly meritocratic and heavily dependent on social networks. Their study indicated that ethnic minorities faced disadvantages not in the formal hiring process itself, but in the preliminary stages before organizational contact. Specifically, minority applicants lacked access to or were less able to utilize the social networks that led to high success rates in securing employment. Other studies also show that low-income neighborhoods often lack the social connections, labor market networks, and other resources essential for economic success (Loury, 2013). Employers may place a high value on these resources, as they are beneficial to their own interests (Darity, 2005).

[9] Women's literacy rates grew from 38% (vs. 46.6% for men) in 2001-02 to 52% (vs. 73% for men) in 2020-21. In 2001-02, only 1.32% of women had a college degree or higher, compared to 2.88% of men. By 2020-21, these numbers rose to 5.5% for women and 6.6% for men (Pakistan Bureau of Statistics, 2022).

Job ads also reflect explicit gender preferences. Matsuda et al. (2019), analyzing job postings on Pakistan's largest job portal, rozee.pk, found that ads mostly favor men for roles requiring higher occupational levels and professional experience. At the same time, women are more often preferred for roles requiring higher education but not for managerial positions.

Researchers argue that patriarchal norms and stereotypes, especially during periods of job scarcity, influence job rationing processes (Braunstein and Heintz, 2008; Azmat and Petrongolo, 2014). Discriminatory hiring practices not only restrict access to better-paying jobs but may also discourage women and marginalized groups from investing in their human capital (Darity and Mason, 1998), perpetuating a cycle of inequality.[12]

On the demand side, prior research in Pakistan has primarily relied on survey data to estimate standard wage regressions for men and women (see Siddiqui et al., 1998, and Sabir and Aftab, 2007). These studies, though limited in number, have used standard Oaxaca-Blinder wage decompositions[13] and quantile regressions to document the persistence of significant gender wage gaps across different economic sectors. Siddiqui et al.(1998), for example, used the Household Income and Expenditure Survey of 1993-94 to decompose the gender-based earning differentials in Pakistan. In their wage decomposition method, they controlled for various characteristics, including age, education, experience, number of working days, area, region, type of work, industrial distribution, employment status in terms of self-employment, and size of the firm. Their results indicated that approximately 55-77 percent of the gender wage gap could not be attributed to observable characteristics. They interpreted this residual as evidence of wage discrimination against women. Sabir and Aftab (2007), using Pakistan Labor Force Surveys from 1996–97 and 2006–07, similarly found persistent wage gaps using both Oaxaca-Blinder decomposition and quantile regressions. Their results show that gender discrimination played an increasing role over time, particularly across the lower and middle portions of the wage distribution.

Despite these early insights, obtaining clear causal evidence of discrimination in Pakistan's labor market remains a challenge (Baert, 2018). Researchers argue that estimates based solely on observational data can often lead to misleading inferences about discrimination due to unobservable differences in workers' productivity (Neumark, 2018; Duflo and Bertrand, 2017; Neumark and Rich, 2019). Workers who appear similar to researchers may seem different to employers, further confounding the analysis (Bertrand and Mullainathan, 2004). Additionally, labor force surveys often lack complete information about workers' characteristics, such as linguistic abilities and academic backgrounds, which can further complicate the understanding of discrimination in the labor market. To address these limitations, researchers have used correspondence testing under resume audits, sending matched fictitious resumes with similar qualifications but differing in social identity signals (e.g., names or addresses) to isolate discrimination in the initial screening process (see Riach and Rich, 2004, for a discussion of the ethical justifications for this methodology).

Another major challenge in Pakistan's labor market is the persistence of low intergenerational income and occupational mobility, with individuals often inheriting their fathers' economic status. Javed and Irfan (2014), for example, found a 72% probability that children of fathers in elementary occupations will enter the same roles. In contrast, only 6.5% of children born in the lowest income quintile reach the highest, with marginally better outcomes for those from middle-income families. While research on this subject remains limited in Pakistan, the cross-country analysis by Grawe (2004)—including data from Pakistan, the U.S., the U.K., Peru, Nepal, Malaysia, and Ecuador—shows that earnings immobility is significantly more pronounced in developing countries. Asadullah (2012) found in rural Bangladesh that 34% of sons from the poorest households remained poor, and 36% of sons from the wealthiest remained wealthy, highlighting greater immobility at the extremes. Globally, studies of intergenerational wealth mobility remain limited (Mazumdar, 2018).

---

[10] In comparison, this figure is 9% in Bangladesh and more than 15% in India. See https://www.statista.com/statistics/1306954/india-women-in-management-positions/

[11] Official statistics might severely underreport women's labor force participation rate as women's unpaid work is not duly accounted for. While women might be employed in productive domains such as domestic farming and handicrafts, cultural and official blinders to their activities represent them as homemakers (see, Ibraz, 1993).

[12] Various supply-side constraints play a significant role in women's labor supply. A significant body of research on women's labor supply in Pakistan has primarily emphasized supply-side dynamics (Shah et al., 1976; Nasir, 2005; Ejaz, 2007; Faridi and Rashid, 2014).

[13] Under this method, to infer gender wage discrimination, separate wage regressions for men and women are estimated. The method decomposes the differentials into two components. One that is captured by differences in characteristics the market associates with productivity (such as education, experience, etc.) and the second that is unaccounted after controlling for all observable characteristics of workers, industry and occupation. The second component of gender disparity in earnings- captured by estimated value of the constant term and coefficients- is interpreted as differential not explained by observable characteristics included in the regression and therefore associated with discrimination.

The broader literature on intergenerational mobility emphasizes the role of residential neighborhoods in shaping individuals' life outcomes and economic opportunities. Studies show that individuals who move to "good" neighborhoods—characterized by lower poverty and more college graduates—experience improved earnings and employment outcomes (Chetty et al., 2014; 2016; 2020). Studies have also highlighted that spatial mobility is linked to intergenerational mobility. "Good" neighborhoods provide more opportunities through better schools, health behaviors, and role models (Sharkey and Faber, 2014; Chetty et al., 2020; Chyn and Katz, 2021).

Theories of spatial mismatch and stratification economics can also explain how segregation limits labor market outcomes. Spatial mismatch literature shows that residential location affects job access, with social networks in segregated areas shaping referrals and employment (Bailey et al., 2020; Krauth, 2004; Fernandez and Mateo, 2006; Schmutte, 2015). Moreover, Elliot (2001) shows that referrals often perpetuate ethnic homogeneity in workplaces. In this context, Mouw (2002) outlines two mechanisms linking residential segregation to employment segregation: physical distance and segregated job networks, both of which restrict opportunities. The spatial-mismatch hypothesis suggests that as jobs move from city centers to suburbs and residential segregation persists, the distance between deprived neighborhoods and areas of job growth increases. The distance between a worker's residence and available jobs increases commuting and job-search costs. This spatial mismatch restricts disadvantaged workers' mobility, contributing to geographic barriers in finding and retaining well-paid jobs (Kain, 1968; Ihlanfeldt, 2006). Employers may prefer candidates who live closer to the workplace, as proximity can enable greater work effort and reliability (Zenou, 2002). Those living farther away may face higher risks of tardiness, absenteeism, and reduced scheduling flexibility (Bunel, L'Horty, and Petit, 2016; Diaz and Salas, 2020; Carlsson and Eriksson, 2023). An alternative perspective suggests that segregated job networks contribute to employment segregation. The socio-demographic makeup of a neighborhood can shape individuals' employment prospects through peer and social network effects—all of which play a significant role in job search processes (Hellerstein et al., 2014; Bunel, L'Horty, and Petit, 2016). Bunel, L'Horty, and Petit highlight that employers often use residential addresses as a screening tool, interpreting neighborhood reputation as a signal of applicant characteristics—a form of "postal code discrimination" also discussed by Zenou (2002). Supporting this, Atkinson and Kintrea (2001) found that 25% to 33% of residents in deprived areas of Scotland believed their neighborhood's reputation negatively affected their employment prospects. Employed individuals were more likely to report stigma than their unemployed counterparts, highlighting the labor market salience of such spatial discrimination. Residents across both deprived areas noted a mismatch between external perceptions and their lived reality, suggesting that neighborhood stigma is less a reflection of actual conditions and more a socially constructed barrier that reinforces exclusion.

These neighborhood-based signaling effects—where employers infer applicants' productive characteristics from their place of residence—are consistent with the theory of statistical discrimination. Phelps (1972) posits that in the face of uncertainty about a worker's future productivity, employers may rely on observable traits such as race, gender, or residential location as proxies for unobservable qualities. Stratification economics further explains that dominant groups sustain inequality to preserve material advantage (Darity and Mason, 1998). Darity (2005), for example, argues that dominant groups derive material benefits from maintaining their privileged position, which incentivizes them to perpetuate discriminatory practices. In market-based economies, such discrimination may persist because discriminatory employers face minimal costs for these behaviors. Arrow (1998) emphasized that social segregation leads to labor market segregation through referral networks, where discrimination may even yield social rewards. Similarly, Thorat and Attewell (2007) argue that residential segregation reflects and reinforces deeper social hierarchies, which labor markets further reproduce through network effects.

It is worth noting that only a limited number of studies have empirically examined employment discrimination based on workers' neighborhoods. Clark and Drinkwater (2000), using data from Britain's Fourth National Survey of Ethnic Minorities, found that individuals residing in areas with high minority concentrations were less likely to be in paid employment, more likely to pursue self-employment, and faced wage penalties—even after controlling for key demographic and job-related factors.

Correspondence studies on this topic remain limited and yield mixed results. Tunstall et al. (2014), in a correspondence study from the UK's labor market, found no evidence that residents of low-income neighborhoods fare worse in low-skilled jobs compared to those in affluent areas. Based on the findings, they argue that "There is no argument for policy interventions, including policies to reduce sorting or to address this neighborhood effect pathway more directly, on the grounds of neighborhood effects on employment." In contrast, Carlsson et al. (2018), in a correspondence testing study from Sweden, showed that applicants with Middle Eastern names from deprived neighborhoods[14] received 42% fewer callbacks than those from affluent neighborhoods. However, they found no neighborhood signaling effect for typical Swedish names.

Overall, while existing literature from developed countries primarily examines how disadvantaged neighborhoods affect workers' labor market outcomes, there is limited evidence on whether neighborhood socioeconomic status influences employers' hiring decisions. This study fills this gap by using a correspondence testing study to analyze how applicants' gender and affluence of residential areas (a proxy for socioeconomic status) affect their likelihood of receiving job interview calls in Karachi's information technology sector.

Karachi's labor market provides a compelling context for this analysis. As one of the world's largest metropolitan areas and the financial capital of Pakistan, it is home to over 18 million people.[15] With a formal economy valued at $114 billion (PPP), Karachi boasts the highest per capita income in Pakistan (World Bank, 2018) and contributes roughly 25% of the national GDP.[16] It is also a major hub for software houses and hosts the largest number of online job postings, with a significant concentration in the IT sector (Bilal et al., 2017). It comprises many ethno-linguistic groups from across Pakistan, allowing a diverse pool of applicants.

Karachi also represents stark social stratification, with access to quality education, healthcare, and infrastructure deeply divided along neighborhood lines, making it a critical site for studying neighborhood signaling effects in hiring processes. In such a context, neighborhood signaling plays a powerful role: an applicant's residential address often acts as a proxy for socioeconomic status, educational background, and perceived professionalism.[17] Affluent neighborhoods provide access to elite English-medium schools, superior infrastructure, and influential social networks—signals of employability and cultural "fit" in professional settings. In contrast, residents of low-income areas face spatial stigma, often associated with disorder, backwardness, or criminality, which can lead employers to view them as risky hires (Hussain et al., 2023; Mallick, 2018). These stereotypes are reinforced by structural disparities in public services, including education, water, and sanitation (Khalil et al., 2023), creating parallel worlds of opportunity. For example, high-income neighborhoods are home to elite English-medium[18] private schools that are often associated with higher educational quality, better infrastructure, and greater social prestige. English-medium schools, largely accessible to affluent families due to high tuition fees and transportation costs, provide direct pathways to reputable universities and high-paid jobs where the English language is widely used.

In contrast, residents of low-income neighborhoods largely rely on underfunded Urdu-medium public schools that often lack adequate resources and qualified teachers. These broader educational deficits create systemic barriers to higher education and professional advancement.[19] This linguistic divide has created two parallel educational systems, reinforcing entrenched class divisions.

[14] They defined "bad" or deprived neighborhoods as those with low socioeconomic status, high immigrant density, and high crime rates, while "good" or affluent neighborhoods featured high socioeconomic status, predominantly native Swedish residents, and a strong safety reputation.
[15] https://worldpopulationreview.com/cities/pakistan/karachi
[16] Dawn News. (2024). Karachi— the neglected golden goose. Retrieved from https://www.dawn.com/news/1844542.
[17] In Pakistan, it is standard practice for job applicants to include their full residential address on resumes, particularly when applying to formal sector jobs via online platforms such as Rozee.pk. This convention is widely followed across sectors and educational levels, and resumes that omit location details are often viewed as incomplete. This practice is reinforced by the application structure on online job portals such as Rozee.pk, which was used in this study. On Rozee.pk, applicants are required to fill in specific fields during the application process, including gender, city of residence, and address. These categories are visible to employers at the initial screening stage.
[18] Pakistan operates a dual educational system with two distinct mediums of instruction that perpetuate socioeconomic stratification. While public schools and many private institutions use Urdu or regional languages (Sindhi, Punjabi, Pashto), elite private schools conduct all instruction in English. Beyond educational quality, affluent neighborhoods also offer access to stronger social networks through socialization.
[19] English proficiency has become increasingly vital for upward mobility in Pakistan's formal labor market.

As Mahboob (2017) argues, the dominance of elite English-medium education—accessible primarily to the wealthy—has created a form of "educational apartheid" that mirrors broader class divides in Pakistan.

Rana et al. (2020) note that Pakistan's urban development follows a core-periphery model that structurally disadvantages peripheral neighborhoods. Exclusionary spatial practices—such as gated communities and the securitization of informal settlements—further entrench these divides and reflect the proto-hegemonic aspirations of a rising middle class (Mallick, 2018).

Moreover, the unequal distribution of public assistance and infrastructure can also contribute to neighborhood stigma, causing residents to experience "dis-utility from low status" (Koster and Van Ommeren, 2023). Being labeled as a 'deprived' neighborhood may result in reputational harm, where the stigma is not only limited to material deprivation but also extends to social cues and stereotypes about the residents themselves. As Besbris et al. (2015) note, individuals living in deprived areas often face suspicion and mistrust in interactions with those outside their neighborhood. This social discrediting can shape how institutions—including employers—respond to residential cues. In stigmatized areas, employers may consciously or unconsciously interpret an applicant's address as a proxy for negative worker traits, such as low productivity, unreliability, or inadequate soft skills.

Karachi, as the country's largest city and a hub of economic activity, offers a sizable pool of job listings and significant variation in neighborhood affluence. The core-periphery divide—with sharp disparities in access to education, infrastructure, and services—makes residential address a strong proxy for socioeconomic status and an ideal site to study neighborhood signaling effects.

## 3. DATA AND METHODOLOGY

### 3.1 Preliminary Insights: Gender Targeting in Job Ads

This study examines employers' hiring practices in Karachi, drawing on job advertisements posted on Pakistan's largest online job portal, rozee.pk. It begins with an analysis of nationwide job advertisements to assess gender-targeted recruitment patterns. Next, a resume audit experiment is conducted, focusing specifically on software programming jobs within Karachi's IT sector. Finally, qualitative surveys with human resource officials provide insights into how employers interpret gender and neighborhood-related signals during the hiring process.

The sectoral and platform choice is motivated by several factors. Software programming jobs are high-skilled, high-paying roles with minimal face-to-face interaction, reducing reliance on subjective "soft skills" (Azmat and Petrongolo, 2014; Banerjee et al., 2009). These jobs are thought to be less prone to discrimination based on gender or neighborhood because they emphasize technical ability. Prior studies (Banerjee et al., 2009) show that discrimination is lower in jobs where productivity is easily observed and interaction is minimal. This study helps to test if that holds in Karachi's IT sector.

IT jobs (especially programming) are frequently posted online, allowing for a large volume of standardized resume submissions necessary for obtaining a reasonable statistical power for the field experiment.

Rozee.pk.[20] is well-suited for this study because (a) many job ads explicitly state gender preferences, enabling an analysis of whether actual callback decisions reflect these biases[21]; (b) the platform allows applicants to track whether their resumes were received and viewed, offering rare visibility into employers' initial screening behavior: and (c) it's the country's largest job portal, widely used by employers in the IT sector, making it a relevant and rich source for sampling real-world hiring behavior.

To gain preliminary insights into explicit gender targeting in job ads, this study extracted data from rozee.pk between October 2018 and July 2019. Job postings during this period were averaged, with Column A representing postings from major cities in Pakistan and Column B focusing specifically on Karachi.

---

[20] According to Raza (2018), Pakistan's IT industry is sized at $2 billion as of 2017-18. It employs more than 12000 employees in its 370 software houses. There are around 300,000 IT professionals while local universities produce around 20,000 IT graduates per year. The data, however, is not disaggregated by gender.

[21] Rozee.pk includes a separate section for preferred gender in its job ads, typically located alongside the required education and experience, which facilitates data collection on advertised gender preferences across different occupational categories.

Table 1 highlights clear gender preferences in job ads across Pakistan. The majority of job ads across Pakistan show a strong preference for men, particularly in fields such as Engineering (29%) and Maintenance/Repair (63%). In contrast, a substantial number of ads (e.g., in Creative Design and Media) reflect a higher percentage of no gender preference. Even in traditionally pink-collar jobs (see Delfino, 2024)—such as those in Health and Medicine—ads favor men (21%) over women (6%). More minor discrepancies are evident in client services and customer support roles, where preferences for men (16.6%) slightly exceed preferences for women (13%). Overall, a substantial proportion of job postings (70%) do not specify any gender preference.

Across job categories in various regions of Pakistan (Column A), male preference for men is lowest in search engine optimization (6.3%), telemarketing (6.5%), media (9.5%), legal affairs (8.7%), executive management (12.5%), and software/web development (13.1%). In contrast, the highest preferences for men appear in categories associated with maintenance and repair (63%), quality control (50%), safety and environment (50%), architecture and construction (41%), computer networking (49%), and database administration (40%). In Karachi (Column B), similar trends persist, with a notably substantially higher male preference for men in maintenance and repair jobs (80%). Nearly 14.5% of software and web development job ads in Karachi express a preference for men, while nearly 84% indicate "no preference." Many IT roles require specific and specialized skills, such as proficiency in programming languages. Due to these unique skill requirements—ads favor men (21%) over women (6%). More pool of qualified applicants may be smaller, making gender less of a consideration. Employers may be more inclined to state "no preference" as long as candidates meet the necessary technical qualifications.  For example, computer networking roles exhibit a stronger preference for men (49% in Column A; 37% in Column B) compared to software programming jobs, which show preferences of 13% and 14.5%, respectively. This discrepancy may arise from the higher degree of face-to-face interaction required in some computer networking positions. Interestingly, women appear to be more favored for teaching positions, likely due to the flexible working hours and alignment with traditional gender roles associated with these professions.

Gender-targeted job ads and varying gender preferences across occupations may reflect paternalistic discrimination, where employers engage in gender-based sorting, penalizing women under the pretense of "protecting" them from perceived harm in roles involving night shifts or long commutes (Buchmann et al., 2024). Buchmann et al.'s framework and field experiments in Dhaka, Bangladesh, demonstrate that this discrimination is driven not by uncertainty about productivity (statistical discrimination) or animus (taste-based discrimination), but by employers' beliefs about workers' perceived welfare.

**Table 1.  Explicit gender targeting of job ads**

| Functional Area | A. Gender preferences in job ads (Pakistan-all areas) | | | | B. Gender preferences in job ads (Karachi) | | | |
|---|---|---|---|---|---|---|---|---|
| | No preference (%) | Men (%) | Women (%) | Total ads | No preference (%) | Men (%) | Women (%) | Total ads |
| Accounts, Finance, and Financial services | 56.0 | 38 | 6.0 | 318 | 54.8 | 40.3 | 4.8 | 124 |
| Administration | 47.1 | 27.1 | 25.8 | 155 | 46.9 | 31.3 | 21.9 | 32 |
| Architects and Construction | 48.1 | 40.7 | 11.1 | 27 | 14.3 | 71.4 | 14.3 | 7 |
| Client Services and Customer Support | 70.4 | 16.6 | 13.0 | 361 | 67.0 | 21.4 | 11.7 | 103 |
| Computer Networking | 51.0 | 49.0 | 0.0 | 49 | 63.2 | 36.8 | 0.0 | 19 |
| Creative Design | 85.5 | 10.5 | 4.0 | 200 | 83.3 | 13.9 | 2.8 | 36 |
| Data Entry | 56.7 | 30.0 | 13.3 | 60 | 60.0 | 26.7 | 13.3 | 15 |
| Database Administration | 60.0 | 40.0 | 0.0 | 10 | 25.0 | 75.0 | 0.0 | 4 |
| Distribution and Logistics | 55.3 | 39.5 | 5.3 | 38 | 53.3 | 40.0 | 6.7 | 15 |
| Engineering | 67.3 | 29.0 | 3.7 | 107 | 58.1 | 41.9 | 0.0 | 31 |
| Executive Management | 75.0 | 12.5 | 12.5 | 8 | 66.7 | 33.3 | 0.0 | 3 |
| Health and Medicine | 73.3 | 21.1 | 5.6 | 90 | 75.9 | 13.8 | 10.3 | 29 |
| Hotel/Restaurant Management | 58.7 | 39.1 | 2.2 | 46 | 41.7 | 58.3 | 0.0 | 12 |
| Human Resources | 62.0 | 22.8 | 15.2 | 92 | 60.0 | 20.0 | 20.0 | 30 |

| Category | | | | | | | |
| --- | --- | --- | --- | --- | --- | --- | --- |
| Import and Export | 46.7 | 43.3 | 10.0 | 30 | 25.0 | 75.0 | 0.0 | 8 |
| Legal Affairs | 87.0 | 8.7 | 4.3 | 23 | 50.0 | 50.0 | 0.0 | 2 |
| Maintenance/Repair | 37.1 | 62.9 | 0.0 | 35 | 20.0 | 80.0 | 0.0 | 10 |
| Manufacturing | 48.9 | 46.8 | 4.3 | 47 | 43.8 | 50.0 | 6.3 | 16 |
| Marketing | 62.1 | 24.3 | 13.6 | 243 | 61.5 | 23.1 | 15.4 | 65 |
| Media-Print and Electronic | 83.3 | 9.5 | 7.1 | 42 | 90.0 | 10.0 | 0.0 | 10 |
| Operations | 50.7 | 38.0 | 11.3 | 71 | 55.6 | 38.9 | 5.6 | 18 |
| Procurement | 59.1 | 31.8 | 9.1 | 22 | 60.0 | 30.0 | 10.0 | 10 |
| Product Development | 64.9 | 24.3 | 10.8 | 37 | 0.0 | 75.0 | 25.0 | 4 |
| Project Management | 69.0 | 17.2 | 13.8 | 29 | 66.7 | 33.3 | 0.0 | 6 |
| Quality Assurance | 73.8 | 21.3 | 4.9 | 61 | 73.7 | 26.3 | 0.0 | 19 |
| Quality Control | 16.7 | 50.0 | 33.3 | 6 | 50.0 | 50.0 | 0.0 | 2 |
| Safety and Environment | 50.0 | 50.0 | 0.0 | 6 | 50.0 | 50.0 | 0.0 | 2 |
| Sales and Business Development | 65.6 | 27.9 | 6.5 | 707 | 58.5 | 33.6 | 7.8 | 217 |
| Search Engine Optimization | 87.4 | 6.3 | 6.3 | 95 | 76.0 | 20.0 | 4.0 | 25 |
| Secretarial, Clerical and Front Office | 46.7 | 45.0 | 8.3 | 60 | 61.1 | 33.3 | 5.6 | 18 |
| Security | 56.3 | 37.5 | 6.3 | 16 | 50.0 | 50.0 | 0.0 | 4 |
| Software and Web Development | 85.5 | 13.1 | 1.3 | 754 | 84.2 | 14.5 | 1.3 | 228 |
| Supply Chain Management | 48.3 | 48.3 | 3.4 | 29 | 50.0 | 50.0 | 0.0 | 4 |
| Teachers/Education, Training and Development | 65.4 | 29.2 | 5.4 | 185 | 61.7 | 6.4 | 31.9 | 47 |
| Telemarketing | 89.7 | 6.5 | 3.9 | 155 | 69.2 | 15.4 | 15.4 | 13 |
| Warehousing | 52.4 | 42.9 | 4.8 | 21 | 60.0 | 40.0 | 0.0 | 5 |

Note: This study extracted data from rozee.pk between October 2018 and July 2019 to examine gender preferences in job ads across Pakistan and Karachi. The total number of job ads is included, along with preferences expressed as percentages. Note that a limited number of job ads indicated preferences for transgender individuals (data not provided in the table); therefore, percentages for gender preferences in some roles may not add up to 100%.

As data on the gender composition of the IT workforce in Pakistan is not readily available through labor force surveys, Table 1 provides insight into the gendered nature of hiring practices. For the IT sector analysis, I classified the following functional areas as IT: Computer Networking, Database Administration, Software and Web Development, and Search Engine Optimization. This classification is based on roles that require primarily technical computing skills and are typically found within IT departments or technology companies. I used a conservative definition focusing on core technical functions to ensure consistency with standard industry categorizations.[22] I excluded borderline categories such as Data Entry, Quality Assurance, and technical Customer Support roles, as these functions, while sometimes IT-adjacent, often exist across multiple industries and may not require specialized technical expertise. Similarly, I did not include Project Management or other roles that might be IT-focused but are not inherently technical in nature. The total IT sector sample comprised 908 job advertisements Pakistan-wide and 276 advertisements in Karachi, representing one of the largest functional categories in the dataset. The IT sector demonstrated the highest rates of gender-neutral job postings, with 83.7% of Pakistan-wide IT advertisements (759 out of 908 total) and 81.2% of Karachi IT advertisements (224 out of 276 total) indicating no gender preference. This high percentage of "no preference" jobs in IT is driven primarily by Software and Web Development (the largest category with 754 ads Pakistan-wide) and Search Engine Optimization, both showing over 80% "no preference" rates, suggesting that IT roles in Pakistan generally demonstrate less explicit gender targeting compared to other functional areas. These findings are consistent with the study by Matsuda et al. (2019), who noted a preference for men in higher occupational levels and with longer professional experience.

Overall, Table 1 suggests that IT jobs are less likely to explicitly mention gender preferences compared to jobs in other sectors. Employers in the IT sector may be less overt in expressing gender preferences in job ads, and it is also possible that gender may be perceived as less relevant to the roles themselves. However, the lack of explicit gender preferences in job ads may not reflect actual hiring practices. Employers might portray themselves as gender-neutral in their advertisements to maintain a positive public image and foster goodwill with clients[23], while still engaging in discriminatory hiring decisions through more subtle methods.

## 3.2 Resume Audit

To examine employers' hiring behavior, this study uses a resume audit methodology, drawing on the experimental designs of Bertrand and Mullainathan (2004) and Galarza and Yamada (2014). Fictitious resumes are constructed to be similar in terms of education, skills, and other productivity-related characteristics, but differ by gendered names (masculine or feminine) and by residential addresses from either high- or low-income neighborhoods in Karachi. These resumes are then submitted in response to software programming job postings in Karachi, and differences in interview callback rates[24] are analyzed. This approach allows for the isolation of discriminatory responses based on gender and socioeconomic indicators of neighborhood.

The research protocol for this study was reviewed and approved by the University of Massachusetts Amherst Institutional Review Board (Protocol ID: 2018-5123). All data collection procedures followed ethical guidelines for digital research involving publicly available information.

The following sections describe the strategy and audit design in detail.

### A. *Applicant's identities: capturing the gender and neighborhood effect*

Signaling gender identity without conflating it with other social identities poses a challenge, as names often carry socio-cultural information related to religion, ethnicity, class, and caste. To address this, this study leveraged the prevalent naming convention found in large cities in Pakistan, including Karachi. Under this naming convention, patronyms are part of the personal name.[25] An individual's name usually consists of two key elements: the masculine and feminine first names and the father or grandfather's name as the last name. For instance, in the name "Hiba Naeem", "Hiba" typically signifies feminine identity while "Naeem" reflects the father's first name, which indicates his gender as a man. For the research methodology, masculine and feminine first and last names were extracted from publicly available student gazette books from schools and universities in Karachi. Names associated with tribes, ethnicity, religious sects, and other social identities were excluded. The most frequently used masculine and feminine first names, paired with neutral last names, are listed in Table A.1 in the Appendix. First names were randomly paired with last names to create combinations for fictitious resumes. Additionally, first and last names were swapped randomly each week.

Identifying high- and low-income neighborhoods for the resume audit is challenging due to ethnic and social segregation in Karachi. To address this, I employed multiple methods. First, I selected neighborhoods with mixed populations based on real estate interviews and surveys. Additionally, I used household income data from the Karachi Household Survey (carried out by Japan International Cooperation Agency (JICA)[26] and cited in the World Bank Report [2018, p.30]) to categorize neighborhoods by household income (see Figure A.1 in the Appendix). Next, I consulted property values on the real estate website, Zameen.com.[27] High-income neighborhoods with higher property values are treated as neighborhoods with high socioeconomic status, and vice versa.[28]

---

[23] IT companies often have global clients in countries where gender equality norms are prioritized.

[24] The outcome variable is a binary variable coded as 1 if the applicant received an interview invitation and 0 otherwise.

[25] This naming convention may be partly influenced by the Arabic language, which has strict gender-related rules that categorize words as feminine or masculine, extending this dichotomy to names (Aribowo, Hadi, and Ma'ruf, 2019). In Pakistan, a predominantly Muslim country where Muslims constitute over 96% of the population, Rahman (2014) notes that women from large cities are more likely to adopt the first names of their husbands or fathers, as a surname, rather than their family names.

[26] For detailed analysis, see the World Bank Report (2018).

In Karachi, high-skilled jobs are predominantly concentrated in the city center, where the high-skilled, affluent population resides. In contrast, low-skilled jobs are scattered around the periphery and near informal settlements (World Bank, 2018). To ensure a balanced selection, I selected high- and low-income neighborhoods that were adjacent and equidistant from the city center. Table A.2 in the Appendix presents the socioeconomic categorization of selected neighborhoods based on average property values and estimated household monthly per capita income. High-income areas such as Clifton, Faisal Cantonment, and Defence Housing Authority (DHA) Karachi exhibit significantly higher property prices and income levels, with average monthly per capita income exceeding PKR 18,000. In contrast, low-income neighborhoods like Korangi and Landhi report much lower income ranges and property values, with per capita income estimates below PKR 6,000. Malir-Model Colony falls in the low-mid category, highlighting the income and housing gradient across the city. This classification captures the spatial concentration of wealth and deprivation in Karachi and serves as the basis for assigning residential signals in the experimental design.

Finally, to control for proximity to job locations and the associated commuting distances, a software app, Distance Calculator, was used to calculate the distance from each applicant's residential address and the advertised job location. Controlling for commute is important, as prior research indicates that applicants from economically disadvantaged areas with limited transportation access may be penalized by employers due to concerns about tardiness, fatigue, or reduced work effort (Diaz and Salas, 2020; Carlsson and Eriksson, 2023). These perceptions can lead to lower callback rates. Controlling for commuting distance also matters to address paternalistic discrimination, wherein employers may restrict opportunities for women in roles involving long (often unsafe) commutes under the pretense of "protecting" them from perceived harm (Buchmann et al., 2024). This helps isolate whether discrimination stems from perceived logistical or welfare concerns or biased assumptions about applicants.

## B. Construction of resumes and the randomization technique

In constructing resumes, I employed a methodology consistent with the approach used by Bertrand and Mullainathan (2004). To ensure that the fictitious applicants resemble actual job market candidates and are competitive, resumes of actual IT professionals posted on LinkedIn.com were consulted. These resumes were then stripped of names, addresses, and any identifiable information, with experience and educational qualifications also altered. After transforming the resumes to prevent them from being traced back to their original owners, the names and addresses of fictitious candidates were randomly assigned.[29] The resumes were periodically rotated among the candidates; for example, a resume assigned to a man from a high-income neighborhood one week was assigned to a woman from a low-income neighborhood the next.

In terms of educational qualifications, all four applicants, who apply simultaneously, share the same institution (University of Karachi) and cohort (a common year of graduation).[30] The University of Karachi (U of K) is the largest public sector university in Pakistan, with an enrollment of 47,337 students and a gender ratio of 60% women to 40% men.[31] Given its scale and the common practice of same-cohort graduates applying around the same time, the risk of employers recognizing resume similarities is low. This mirrors typical job search patterns and supports the credibility of simultaneous applications—a strategy consistent with other correspondence audit studies, such as Galarza and Yamada (2014), who also submitted four equivalent resumes per job ad.

---

[27] Conversations with real estate agents in Karachi provided valuable insights into property values across different neighborhoods. Maps of neighborhoods, some available from www.zameen.com and others provided by the agents, helped generate a list of residential addresses for this experiment. Additionally, Zameen.com publishes property blogs discussing the socioeconomic status of various neighborhoods in Karachi, which were instrumental in selecting a combination of high- and low-income areas with diverse populations.

[28] See Thorat and Attewel (2007) for a Weberian notion of status groups. In the Indian context, Vakulabharanam's (2010) class schema identifies two classes in urban areas: elite (owners, managers, and professionals) and the workers (other than professionals).

[29] While gender and neighborhood were fully randomized, distance was not randomized conditional on gender and neighborhood. Instead, once residential addresses were assigned to resumes (as part of the neighborhood treatment), the corresponding commute distances were calculated from those assigned addresses to the job locations

[30] Maintaining similarity in educational qualifications within Karachi's context is challenging, as educational background signals important aspects of social class and privilege. A significant educational divide exists, with high-quality, expensive, private English-medium institutions (typically from nursery to high school) available only to a small elite group, which is socially regarded as a marker of power and privilege (see Mahboob, 2017). By omitting prior schooling from the resumes and ensuring that the bachelor's degree institution is consistent, the goal is to minimize potential confounding factors.

[31] https://www.timeshighereducation.com/world-university-rankings/university-karachi

The University of Karachi is known for its diverse student body and a well-regarded computer science program.[32] Ranked 58th in South Asia[33], it provides accessible education to students from lower-income backgrounds, while its computer science program's reputation also attracts applications from more affluent students. This socioeconomic diversity helps ensure that the applicants appear plausible across a range of class identities. Using the same institutional affiliation allows for greater control over resume content while also accounting for employers' potential class-based assumptions about alma mater.

### C. Collection of data: recording interview callbacks

Each fictitious application was assigned a unique phone number and email address to track interview callbacks. Since voicemail was rarely used at the time, employers typically relied on phone calls. Trained student researchers answered these calls, recorded details in a spreadsheet, and informed callers that the candidate was no longer job seeking, having accepted another position, an approach consistent with standard audit study protocols. This methodology follows established ethical guidelines for such field experiments, where researchers must balance the need to document discrimination against potential harm to study participants (Riach and Rich, 2004).

### D. Empirical methodology

To estimate the effect of gender and neighborhood on callbacks for interviews following specification is used:

$$D_i = \alpha_0 + \alpha_1 \textbf{Gender}_i + \alpha_2 \textbf{Neighborood}_i + \alpha_3 \textbf{Distance}_i + \alpha_4 \textbf{FirmSize}_i + \alpha_3 \textbf{Sector}_i + \varepsilon_i$$

Where $D_i$ is the callback dummy and takes the value 1 if applicant i gets a callback for a job interview, and 0, otherwise. *Gender* and *Neighborhoods* are dummy variables, and coefficients $\alpha_1$ and $\alpha_2$ are the parameters of interest for this study. $\textbf{Distance}_i$ is the distance between the residential address of job applicant i and the location of the advertised job. To address unobserved heterogeneity, the specification controls for firm size (total employees) and sector, factors particularly relevant given evidence that IT and engineering firms often exhibit male-dominated environments (Koput and Gutek, 2010).

Based on preliminary insights from job ad data and secondary surveys of Pakistan's job market, two hypotheses emerge. First, because software programming roles typically involve less face-to-face interaction than sectors like sales or business development, gender differences in interview callbacks may be minimal, especially since the study focuses on IT employers who do not explicitly state any gender preferences in job ads. This aligns with Fernandez and Campero (2017), who analyzed over 250,000 job applications across 441 tech firms in California and New York and found limited evidence of gender-based screening biases, regardless of organizational hierarchy. Second, if residential neighborhoods convey information about expected productivity (Carlsson and Eriksson, 2023) or social capital (Koput and Gutek, 2010), we may observe substantial variation in callbacks based on neighborhood characteristics. These hypotheses also draw on insights from Combes et al. (2016), who show that areas with higher concentrations of French natives exhibit larger African-French employment gaps, with discrimination effects amplified in regions with more customer-contact positions. A one standard deviation increase in French native concentration widens the unemployment gap by 24-37% and the contact job gap by 13% of their respective standard deviations.

---

[32] https://www.scimagoir.com/rankings.php?sector=Higher+educ.&country=PAK&area=1700, and
https://www.universityrankings.ch/results/QS/2024?ranking=QS&year=2024®ion=&q=Pakistan
[33] https://www.topuniversities.com/universities/university-karachi

### 3.3 Qualitative Survey of Human Resource (HR) Officials

This study also conducted a survey of HR officials to understand how employers interpret gender and neighborhood information during the screening process. Given that the surveys were conducted online, were anonymous, voluntary, and offered no incentives for participation, it was essential to cast a wide net to counteract the high non-response rate. Invitations to participate in the survey were sent to actual HR officials in two major cities: Lahore and Karachi.[34] To further reduce any potential effects of social desirability bias, data on personal identifiers, including the gender of the respondent, were not collected. The survey was conducted from October 2018 to July 2019 and focused on human resource officials in Information Technology (IT) firms and Sales and Business Development (SBD) firms for multiple reasons. Firstly, in IT jobs, face-to-face interactions with customers and colleagues are often limited, suggesting that any biases or discriminatory preferences may be more closely related to employer practices. In contrast, SBD positions typically require greater direct interaction with customers and colleagues. Additionally, IT roles generally involve more remote work and less travel than SBD positions. Data from rozee.pk shows that Sales and Business Development roles are more male-dominated than IT jobs in Pakistan (see Table 1 and Matsuda et al., 2019). This comparison provides valuable insights into the complexities of discrimination across different sectors.

The survey also explored how gender and socioeconomic status influence perceived productivity—perceptions that may differ by sector based on labor force composition and role expectations. By identifying how stereotypes vary across sectors, the study uncovers how discriminatory attitudes operate beyond explicit job ad preferences, and how social signals are interpreted differently across sectors.

The study collected a total of 201 completed questionnaires, with 101 responses from professionals in IT and 100 from those in SBD firms. The survey included both closed- and open-ended questions. Two key inquiries were:

1) Do you think men and women are similar in work characteristics (e.g., productivity, punctuality, and professionalism)?
2) Do you think job market candidates from high-income areas and low-income areas are similar in work characteristics (e.g., productivity, punctuality, and professionalism)?
The questionnaire also included follow-up probing questions such as:

- What kind of work characteristics do you associate with men and women?
- What kind of work characteristics do you associate with candidates/workers from high-income areas?
- What kind of work characteristics do you associate with candidates/workers from low-income areas?

These qualitative surveys provide insight into employers' stated beliefs. However, to assess employers' actual hiring behavior, the study uses a resume audit experiment, which tests whether hiring discrimination based on gender and neighborhood affluence persists when such biases are not explicitly stated.

## 4. RESULTS

### 4.1 Resume Audit Results

A total of 2032 resumes were submitted in response to 508 online job ads with no explicit gender preferences—1016 for high-income neighborhoods and 1016 for low-income neighborhoods, evenly split between names corresponding to men and women within each group. The resume audit spanned 13 months, from September 2018 to October 2019. Job applications did not require the submission of photographs, helping to minimize confounding effects related to physical appearance. Prior research suggests that physiognomy and beauty can influence perceived favorability and chances of being hired (Hamermesh and Biddle, 1993; Mobius and Rosenblat, 2006; Doorley and Sierminska, 2015). Resume audit studies have also found that perceived beauty, manipulated through applicants' photos, significantly affects callback rates for job interviews (Galarza and Yamada, 2014; López Bóo et al., 2013; Deng et al., 2019). To isolate the effects of gender and neighborhood affluence, this study excluded photographs to avoid potential appearance-based biases.

---

[34] The survey aimed to gather a sufficiently large and diverse pool of HR respondents, and conducting it in Karachi and Lahore—two major metropolitan cities—was crucial for this goal. These cities are home to numerous business centers, making them ideal locations for reaching a diverse pool of HR professionals.

Table 2 presents callback rates for job interviews by gender and neighborhood. Parentheses below the callback rates show the number of resumes sent, while column 4 reports p-values from tests of equal proportions. For all resumes combined, the difference in callback rates between high-income (8.27%) and low-income (5.71%) neighborhoods appears statistically significant (p-value < 0.05), representing a 2.56 percentage point advantage(or neighborhood premium) for resumes from high-income areas. The callback rate is 45% higher for candidates in high-income neighborhoods (ratio = 1.45), indicating a meaningful neighborhood-based hiring disparity.

When analyzed by gender, men show a marginally significant neighborhood effect (p-value < 0.10) with a 3.15 percentage point difference in callback rates between high-income (9.45%) and low-income (6.30%) neighborhoods. For women, while callback rates follow the same pattern with higher rates in high-income neighborhoods (7.09% vs. 5.12%), the difference (1.97 percentage points) doesn't appear statistically significant at the conventional level of significance (p-value > 0.10).

Rows 2 and 3 display callback rates based on gender for each neighborhood. Women from high-income neighborhoods receive a 7.09% callback rate compared to 5.12% for those from low-income areas. This difference, however, is not statistically significant (p-value > 0.050). Men from high-income neighborhoods have a 9.45% callback rate versus 6.30% for men from low-income areas, representing a premium that is marginally significant at the 10% level of significance.

Analyzing the mean callback rates by gender (Table A.3 in Appendix) reveals a substantial disparity: 7.87% for men versus 6.10% for women, reflecting an observed difference of 1.77 percentage points—nearly a 29% higher callback rate for men compared to women. However, the minimum detectable effect (MDE) for this sample size is about 3.16 percentage points (see Appendix), meaning differences smaller than this cannot be reliably detected with 80% power. The analysis underscores the importance of considering practical significance when interpreting results from resume audit studies. The resume audit process also found that jobs with an explicit preference for men block applications from women at the submission stage, regardless of qualifications. This practice highlights systemic biases in hiring processes that significantly disadvantage qualified women in the job market. Since a substantial portion of the gender penalty against women is likely represented through gendered job ads, gender differences in callback rates for job ads where gender preferences are not explicitly stated could indicate an additional and significant element of implicit discrimination.

## Table 2. Mean callback rates by gender and neighborhood

|  | Percent callback for high-income neighborhoods | Percent callback for low-income neighborhoods | Ratio | Percentage difference (p-value) |
|---|---|---|---|---|
| All resumes | 8.27 (1016) | 5.71 (1016) | 1.45 | 2.56 (0.024) |
| Women | 7.09 (508) | 5.12 (508) | 1.38 | 1.97 (0.190) |
| Men | 9.45 (508) | 6.30 (508) | 1.50 | 3.15 (0.062) |

Notes: Callback rates for the entire sample and subsamples for high-income neighborhoods (column 1) and low-income neighborhoods (column 2). The number of resumes sent in each sample and subsample is in parentheses. Column 4 shows the p-value for a test of proportion with the null hypothesis of equal callback rates across groups.

According to Matsuda et al. (2019), who analyzed job posting data on rozee.pk, skill demands from employers vary by gender. Jobs advertising a gender preference for women are less likely to require programming skills (13%). Conversely, such skills are more likely to be required in jobs that prefer men (50%). Jobs indicating a preference for women are also more likely than those advertising a preference for men or equal employment opportunity to require expertise in communication, Microsoft Office, and other soft skills (e.g., linguistic abilities)." This suggests that while gendered job ads play a significant role in gender-based differences in callback rates, there may be additional, subtler forms of discrimination or biases at work that continue to affect the hiring process and contribute to gender disparities in callbacks, even in cases where job ads do not explicitly specify gender preferences.

Table 3 examines the effect of gender and neighborhood on callbacks for job interviews, including controls and an interaction term.[35] Consistent with similar resume audit studies, such as Glaraza and Yamada (2014), the table includes controls for firm size (measured by the number of employees) and the secondary economic sector of activity.[36] See Table A.6 in the Appendix for mean callbacks by firm size and Table A.7, which categorizes economic sectors of activity under the ILO's International Standard Industrial Classification of economic activities. Controlling for commute distance is essential for investigating the neighborhood signaling effects (Carlsson et al., 2018; Diaz and Salas, 2020; Phillips, 2020; Carlsson and Eriksson, 2023). However, disentangling the impact of commuting from other potential neighborhood effects poses a challenge. Commute times through public transit can significantly fluctuate due to traffic congestion in different areas of Karachi. Hence, this study uses the distance from a candidate's address to the advertised job location (measured in kilometers) as a proxy for commuting distance.

### Table 3. Effect of gender and neighborhood on callback rate: OLS regression

| | (1) | (2) | (3) |
|---|---|---|---|
| Gender (1= Man, 0= Woman) | 0.018 | 0.027* | 0.022 |
| | (0.011) | (0.015) | (0.021) |
| Neighborhood (1= high-income, 0= low-income) | 0.026** | 0.032** | 0.030 |
| | (0.011) | (0.015) | (0.02) |
| Gender × Neighborhood | | | 0.011 |
| | | | (0.03) |
| Distance | | -0.004** | -0.004** |
| | | (0.001) | (0.001) |
| Distance × Gender | | | |
| Distance × Neighborhood | | | |
| Control for firm size | No | Yes | Yes |
| Control for economic sector of activity | No | Yes | Yes |
| Constant | 0.048*** | 0.21*** | 0.21*** |
| | (0.01) | (0.05) | (0.05) |
| Mean of Dep. Var. | 0.069 | 0.077 | 0.077 |
| Observations | 2032 | 1236 | 1236 |
| R-squared | 0.004 | 0.017 | 0.017 |

Note: Robust standard errors are in parentheses: *** p<0.01, ** p<0.05, * p<0.1

In Table 3, column 1, the coefficient on the neighborhood is 0.026 and is significant at the 5% level, indicating that residing in a high-income neighborhood increases callback probability by 2.6 percentage points. In column 2, the coefficient increases slightly to 0.032 and remains significant at the 5% level. Distance has a negative and significant effect: each additional kilometer reduces the callback rate by 0.4 percentage points (or 0.64 percentage points per mile[37]), suggesting employers may penalize applicants with longer commutes. Gender effects appear more nuanced based on the addition of control variables and interaction effects. In Column 3, the interaction between gender (1 = Man) and neighborhood (1 = high-income) is positive but statistically insignificant. This suggests that while each variable (being a man and being from a high-income neighborhood) has its effect on the callback rate, the combined effect does not significantly differ from the sum of their individual effects. In other words, the additional benefit of being both a man and from a high-income neighborhood does not considerably change the callback rate beyond what is expected from each variable individually. One possible explanation for this pattern is that the study focuses specifically on employers who did not explicitly state gender preferences in their job ads. Since explicit gender preferences are partially captured through gender-targeted advertisements, any observed discrimination based on an applicant's neighborhood may reflect additional biases not accounted for in these stated preferences.

---

[35] To address concerns about missing observations, Table A.4 in the Appendix reports regressions on the restricted sample (excluding missing controls) without covariates. The results remain consistent, reinforcing the robustness of the distance and neighborhood effects.

[36] While most companies in this study primarily operate in the IT sector, some also have secondary operations in other sectors. While the number of observations in sectors outside of IT is relatively low, including these sectors allows to control for any residual sector-specific effects that might affect the main analysis.

The positive and statistically significant coefficients for high-income neighborhoods suggest that employers favor applicants from more affluent areas, even after controlling for commuting distance, firm size, and sector. Two explanations, among others, may be particularly relevant in Pakistan's context. First, theoretical arguments related to statistical discrimination theory suggest that there could be increased uncertainty regarding the unobserved productive traits of job applicants from low-income neighborhoods (Carlsson et al., 2018). This uncertainty could be particularly pronounced due to assumptions about early educational quality, as resumes do not include information about primary and secondary schooling, signaling only the applicant's bachelor's degree institution and performance. On the other hand, employers may associate elite neighborhoods like Clifton or DHA Karachi with higher productivity, professionalism, or educational quality, due to better infrastructure, schooling, and exposure to English-medium instruction. At the same time, applicants from low-income areas—such as Korangi or Landhi—may face stigma or stereotypes based on perceptions of criminality or unreliability.

Second, neighborhoods with high social prestige, status, and influence may signal superior access to professional networks and social capital—resources that employers strategically value (Hellerstein et al., 2014; Bunel et al., 2016; Erikson, 2017). This preference reflects what Fernandez, Castilla, and Moore (2000) identify as employers acting as "social capitalists," viewing workers' social connections as investments that yield economic returns through three key mechanisms: accessing richer talent pools, achieving better job-candidate matches, and enabling social enrichment of the organization. High-income neighborhoods typically harbor extensive networks of influential professionals, business leaders, and educated elites who can provide strategic connections, client referrals, and industry insights. Employers may view hiring candidates from affluent areas like Clifton or DHA Karachi as an opportunity to tap into these embedded social resources, especially when faced with a scenario of identical applicants from low-income areas.

Overall, the results in this section empirically examine two key aspects of discrimination to understand how socioeconomic segregation may influence employers' hiring practices: neighborhood signaling effects (captured through addresses in high- and low-income neighborhoods) and spatial mismatch (captured via commuting distance between an applicant's residential location and job location). It's important to identify whether employers discriminate based on the socioeconomic reputation of a neighborhood, known as neighborhood signaling (Bunel et al., 2016; Carlsson and Eriksson, 2023), or based on the commuting burden associated with long travel times, referred to as spatial mismatch (Zenou, 2002). Distinguishing between these mechanisms is critical for identifying appropriate policy interventions: whether to focus on reducing informational stigma attached to certain areas or on improving transportation and mobility for job seekers. In this study, applicants from high-income neighborhoods received callbacks at a rate of 8.27%, compared to 5.71% for those from low-income areas—a statistically significant difference of 2.56 percentage points (p = 0.024), or a 45% relative advantage (Table 2). Regression analysis confirms that this neighborhood premium remains robust even after controlling for commute distance, firm size, and industry (Table 3). On spatial mismatch related to commuting distance, results suggest that a 5-kilometer (3.11-mile) increase in distance results in a 2-percentage-point decline in the likelihood of being contacted for an interview. These findings align with Diaz and Salas (2020) in Colombia and Carlsson and Eriksson (2023) in Sweden, both of whom find that commuting distance significantly reduces callback rates. However, unlike those studies, the results here suggest that both neighborhood signaling and spatial mismatch are at play: employers in Karachi not only penalize applicants with longer commutes but also respond to the stigmatized reputation of low-income neighborhoods, potentially tied to assumptions about reliability, education quality, or crime. This dual structure of discrimination mirrors findings from Bunel et al. (2016), who report a 9.4 percentage point callback gap in France for applicants from disadvantaged neighborhoods—even when distance is held constant. Their study, which sent 2,988 fictitious applications to 498 job offers in the Paris metropolitan region (targeting waiter and cook roles), shows that both neighborhood and broader administrative reputation shape hiring outcomes. Notably, they also identify a distance penalty for jobs located in Seine-Saint-Denis, indicating that commuting proximity does not neutralize discrimination; employers may still penalize applicants from nearby but stigmatized areas. The magnitude of the neighborhood signaling effect in Karachi is similarly substantial and persistent. Gender disparities are also evident, though more nuanced: men received nearly 1.8 percentage points higher callbacks than women (7.87% vs. 6.10%), but this difference falls below the minimum detectable effect size given the sample. In Table 3, Column 2, the most conservative specification with controls for firm size, sector, and commute distance, the gender coefficient is positive and marginally significant at the 10% level ($\beta = 0.028$, p = 0.065). These findings are directionally consistent with Zhang et al. (2021), who document a 35% or 7.6 percentage point lower callback rate for women in China's graduate job market.

---

[37] A 5-mile increase in commuting distance is associated with a 3.2 percentage point decrease in the likelihood of receiving a callback.

Table A.5 in the Appendix explores the interaction effects. The estimated effects of gender, neighborhood, and their interactions with distance appear stable in sign, magnitude, and statistical significance across different model specifications. Men are substantially more likely than women to receive a callback for a job interview. Applicants from high-income neighborhoods also have a significantly higher chance of receiving callbacks. Interestingly, gender and commuting distance interact in complex ways, and failing to control for these can mask the actual gender effect. The Gender × Distance interaction term is negative and significant, suggesting that men are penalized more for longer commutes than women. One possible explanation is that employers may hold higher expectations for men to be mobile, punctual, and consistently available. In Karachi's urban context, long commutes for women are generally accepted, especially due to the presence of gender-segregated buses and commuter vans. Thus, employers may interpret women's long commutes differently—for example, assuming greater job commitment if they are willing to apply to jobs farther from their residence. While paternalistic discrimination typically restricts women on the assumption that they need protection from risky or demanding conditions (e.g., unsafe commutes), this same framework may lead employers to view women's distance as an understandable constraint, linked to family responsibilities or mobility limitations, and not as a reflection of low commitment or professionalism.

It is also noteworthy that the gender coefficient increases in magnitude when the Gender × Distance interaction is included, reflecting variation in how commuting distance affects men and women. In models without this interaction, the estimated gender effect averages across all distances, potentially obscuring important heterogeneity in employer behavior. Specifically, the male advantage in callbacks for applicants living close to the job may be understated, as the model fails to account for the steeper penalties men face with longer commutes. Including the interaction allows the effect of distance to vary by gender, revealing that men receive significantly more callbacks at shorter distances but are penalized more sharply as distance increases.

These findings underscore the need to examine discrimination across different occupational contexts. My audit study focused on Karachi's IT sector, where technical skills are prioritized. Still, roles often involve elements of client interaction and communication—particularly when software firms engage with global clients, which requires English proficiency. As a result, employers may screen applicants based on both perceived professionalism and social status, which are often taken as proxies for linguistic ability. A relevant comparison comes from Banerjee et al. (2009), who conducted a large-scale resume audit in Delhi's software and call-center sectors to examine caste-based hiring discrimination. While no significant differences in callback rates were found between upper-caste and Other Backward Caste (OBC) applicants for software jobs (4.64% vs. 4.28%), the disparity was more pronounced in call-center jobs, where upper-caste applicants received callbacks at a rate of 20.81% compared to 13.02% for OBCs—a 7.79 percentage point gap and a 60% relative advantage. The findings suggest that caste discrimination is more likely in occupations emphasizing soft skills, where employers may rely on social background cues such as surnames to infer suitability.

Altogether, results in this section suggest that discriminatory hiring in Karachi's IT sector is not solely driven by explicit gendered job ads, but also by subtle, layered forms of bias tied to geography, gender, and perceived employability.

### 4.2 Qualitative Survey Results

Figure 1 displays the main results of the survey. In the first column of the figure, a comparison is drawn between the attitudes of HR professionals in IT and Sales and Business Development (SBD) firms regarding the associations between neighborhoods and work characteristics. The second column compares attitudes toward men and women. In the Information Technology (IT) sector, 50% of respondents indicated that men and women share similar work characteristics (in terms of productivity, punctuality, and professionalism). In SBD firms, 70% of HR professionals believe men and women differ in productivity, punctuality, and professionalism.

**Figure 1. Attitudes toward candidates based on gender and neighborhood**

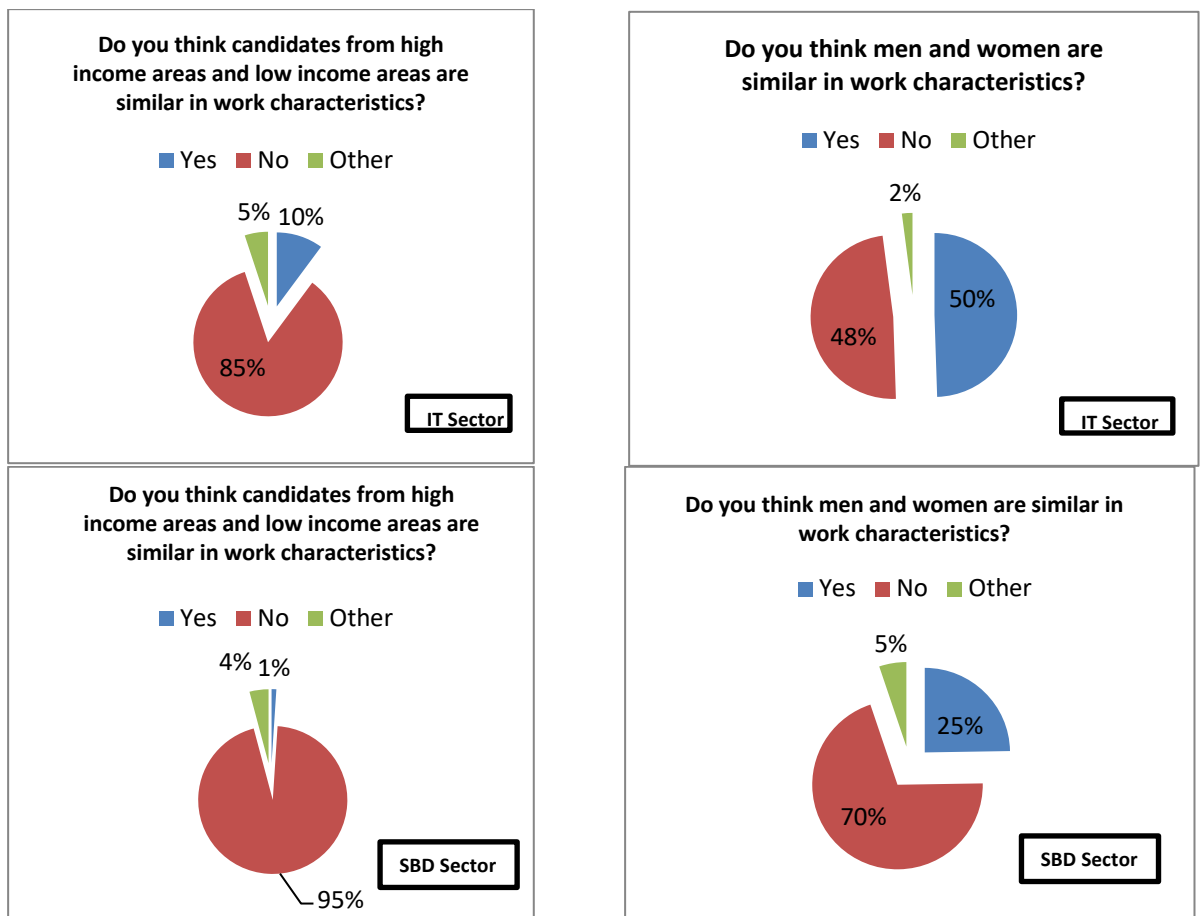

Note: Results of the survey data collected over October 2018-July 2019.

In the second column, 85% of HR professionals in IT firms and 95% in SBD firms state that a candidate's residential location significantly affects perceived work characteristics like punctuality, productivity, and professionalism. These views suggest that employers may use residential location—and by extension, commuting distance and socioeconomic status—as signals of worker reliability or commitment. Such biases are especially influential when employers face uncertainty about applicants' unobservable traits or hold deeply rooted stereotypes. Differences in HR attitudes across IT and SBD firms may reflect the nature of the roles: IT jobs often allow remote work and flexible hours, whereas SBD roles require frequent commuting and face-to-face interaction with clients and colleagues. As a result, both gender and residential location may carry greater weight in hiring decisions within SBD firms. These views suggest that employers may use residential location—and by extension, commuting distance and socioeconomic status—as signals of worker productivity, punctuality, and professionalism. This pattern also aligns with the findings from Buchmann et al. (2024), which suggest that employers may exclude women from certain roles due to concerns about night shifts, unsafe commutes, or ingrained assumptions about gendered suitability.

In their open-ended responses, most participants viewed men and women as generally comparable in terms of education and skills. However, women were more often associated with roles requiring care and patience, while men were linked to jobs emphasizing punctuality, competitiveness, and physical labor. This indicates that employers may weigh gendered traits—beyond qualifications—when assessing candidates during the hiring process. It's noteworthy that in their open-ended responses, most participants saw men and women as generally similar in education and skills. However, women were often linked to jobs requiring care and patience, while men were associated with roles emphasizing punctuality, competitiveness, and intensive labor. This suggests that employers may consider additional characteristics beyond education and skills when evaluating candidates during the hiring process.

**Table 4. Textual analysis from open-ended responses: Gender and neighborhood perceptions in hiring**

| Category | Content analysis of survey responses |
|---|---|
| Gender Differences | Men and Women Generally Similar |
| Associations Between Work Characteristics and Neighborhoods | |
| General Associations | Low Income: Hardworking, Interactive, Less Professional<br>High-Income: Skilled, Educated, Suitable for Corporate Culture |
| Punctuality vs. Professionalism | Low-Income: More Punctual but Less Professional<br>High-Income: More Professional but Less Punctual |
| Education-Related Associations | Low-Income: Less Educated Individuals<br>High-Income: Higher Quality of Education |
| Job Mobility and Commitment | High-Income: Less Likely to Stay Long-Term |
| Field vs. Desk Job Preferences | Low-Income: More Comfortable with Field Jobs<br>High-Income: Prefer Desk Jobs |
| Communication Skills | Low-Income: Poor Communication Skills<br>High-Income: Good Communication Skills |
| Leadership Characteristics | High-Income: Exhibit More Leadership Characteristics |
| Adaptability | Low-Income: More Adaptable |

A thematic analysis of the open-ended responses revealed distinct perceptions of candidates from low- and high-income areas. To conduct this analysis, NVivo and Excel were used to identify recurring themes and relationships within the text. Responses were coded into key categories—such as punctuality, professionalism, education, and job commitment—to enable structured comparisons across gender and neighborhood indicators. NVivo's query functions and word frequency tools helped quantify the prevalence of specific perceptions in open-ended responses (e.g., "men and women are generally similar") and supported the generation of visualizations such as coding matrices and word clouds, which informed the thematic interpretations summarized in Table 4.[38] Results show that respondents often described individuals from low-income areas as hardworking but also linked them with negative traits such as limited skills, lower productivity, and a lack of professionalism. In contrast, candidates from high-income areas were generally perceived as qualified, skilled, well-educated, and hardworking, though some were seen as less punctual.

---

[38] The word frequency queries counted and ranked the most commonly used words or phrases in open-ended responses. It also helped in uncovering dominant themes and repeated language HR officials uses.

Overall, respondents from both the IT and SBD sectors shared similar perceptions regarding work attributes associated with candidates from high- and low-income neighborhoods. Respondents across both IT and SBD sectors expressed similar views about candidates from high- and low-income neighborhoods. High-income candidates were often seen as less likely to stay in the same job long-term, more inclined toward desk jobs, and characterized by strong communication, leadership, and adaptability. These perceptions reflect how hiring decisions may be shaped by both job requirements and implicit biases tied to socioeconomic background (signaled through neighborhood affluence).

## 5. CONCLUSION

This paper investigates whether, and in what ways, employers in Karachi's information technology sector screen candidates based on observable identity signals such as gender and neighborhood affluence—a proxy for socioeconomic status. Using a three-pronged research strategy—analysis of job advertisements, a randomized resume audit experiment, and qualitative surveys of human resource professionals—the study finds consistent evidence of discriminatory practices.

The job ad analysis reveals widespread gender-targeted language that limits opportunities for women before the interview stage. The data shows a general preference for men across most occupations, even in traditionally pink-collar roles.

The second part of the study examines hiring behavior among employers who do not indicate gender preferences in job ads by conducting a resume audit. A total of 2,032 resumes were submitted in response to 508 full-time job openings with no specified preferences about gender and residential location. The results reveal a significant disparity in callback rates between candidates from high- and low-income neighborhoods, with a 45% higher rate for those from affluent areas. This "neighborhood premium" remains significant even after controlling for commuting distance. After controlling for neighborhood, firm size, economic sector, and commuting distance, men receive 2.8 percentage points more callbacks than women. Candidates from high-income neighborhoods receive 3.2 percentage points more callbacks than those from low-income areas, holding gender and other controls constant. Additionally, the interaction between gender and neighborhood is not statistically significant, suggesting that the combined effect of being a man and from a high-income neighborhood does not exceed the individual effects of each variable. Interestingly, the Gender × Distance interaction term is negative and statistically significant, suggesting that men are penalized more for longer commutes than women. One possible explanation is that employers may expect men to be more mobile and punctual, while interpreting women's willingness to undertake long commutes as a potential signal of job commitment and professionalism—reflecting how paternalistic discrimination may play a nuanced role in Pakistan's context.

Given that this study focuses on employers in Pakistan's IT sector who do not state gender preferences, it is likely that women in other sectors face even greater discrimination, as reflected in explicitly gendered job ads. Since expressing gender preferences in ads is permitted, discrimination is embedded in the hiring process from the outset.

Moreover, favorable treatment of candidates from affluent neighborhoods may be driven by two main factors: first, employers may view residential location as a proxy for productivity, especially where neighborhoods differ significantly in access to education, healthcare, and other resources. Second, the neighborhood may reflect the values and social capital a candidate brings, which may align with the employer's social networks, particularly if the employer resides in a similar area.

To further understand hiring attitudes, the study conducted a double-blind qualitative survey among HR professionals in IT and Sales and Business Development (SBD) firms. The survey included open-ended questions about beliefs regarding the similarities in work attributes between men and women and between candidates from different socioeconomic backgrounds. Findings show a stark contrast between sectors: 48% of IT and nearly 70% of SBD respondents believe men and women differ in work traits. A majority of respondents (85%) believe a candidate's place of residence plays a key role in shaping their productivity, punctuality, and professionalism—this belief is even more pronounced in the SBD sector (95%). Overall, more employers associate positive work attributes with candidates from high-income neighborhoods.

It is important to acknowledge some limitations of the study. While Karachi is Pakistan's largest metropolitan and financial hub with a significant IT sector, the study focuses on a specific segment of the labor market and may not fully reflect the broader labor market. Additionally, not all IT employers advertise job openings online or on the portal used in this study, though previous analysis suggests IT jobs are more often posted online. Some use alternative methods like referral networks or university and alumni job boards. These varied practices may lead to under- or overreporting of employment discrimination. Finally, while callbacks are useful indicators of early-stage bias, they do not fully capture actual hiring outcomes, which could reflect different levels of discrimination after multiple interview stages.

Despite these limitations, this study carries significant implications for both local and global contexts. The pronounced neighborhood penalty against candidates from low-income areas may have lasting effects, potentially discouraging individuals from these neighborhoods from pursuing further education and skill development. This, in turn, could perpetuate and deepen social inequalities in educational attainment and labor market opportunities. Moreover, the study provides a broader perspective on employment discrimination, suggesting that penalties linked to residential neighborhoods may also reflect limited intergenerational income and occupational mobility for low-income groups.

Overall, the findings highlight the presence of both explicit and subtle forms of employment discrimination, emphasizing the urgent need to reform and strengthen labor market policies in Pakistan. Without such measures, gender-targeted job ads and discriminatory hiring practices will continue to hinder equal opportunities for women and candidates from low-income communities.

**Acknowledgments:** I would like to express my gratitude to the participants of the 2023 and 2024 International Behavioral and Public Policy Conferences at the University of North Carolina at Chapel Hill and the University of Cambridge, UK, as well as the 2020 Advances with Field Experiments Conference at the University of Chicago, who provided valuable feedback on earlier drafts of this paper. I am especially thankful to the esteemed faculty at the University of Washington and the University of Massachusetts Amherst. I also appreciate the comments, suggestions, and questions provided by the anonymous reviewers, whose thoughtful insights significantly enriched the discussion and analysis presented in this paper. This research received generous support through grants from the US Fulbright Program and the Political Economy Research Institute at the University of Massachusetts Amherst. Special thanks to the undergraduate student research assistants at Lahore University of Management Sciences, Pakistan, for their excellent research support. Noman Khattak and Raza Mustafa played key roles in conducting fieldwork and providing research assistance.

**Additional Notes and Data Availability:** The study was approved by the Institutional Review Board (IRB) at the University of Massachusetts Amherst. This approval was granted after a thorough review of the study's design, including the use of fictitious applications. The IRB's approval ensures that the study adheres to ethical guidelines and considers the welfare of the participants involved. All the data, therefore, remains confidential. The anonymized dataset, description of the tools, the survey questionnaire, and other supplementary materials are available upon request.

**Appendix**

**Table A.1. List of names for resume audit**

| Feminine first names | Masculine first names | Last names |
| --- | --- | --- |
| Hiba | Ahmed | Iqbal |
| Javeria | Naeem | Zahid |
| Saba | Tariq | Atif |
| Zara | Jamal | Ahmed |
| Sara | Najeeb | Mehmood |
| Saba | Nabeel | Mustafa |
| Hira | Khalid | Yosuf |
| Javeria | Asif | Iqbal |
| Saima | Rehan | Imran |
| Uzma | Fahad | Tariq |
| Naila | Ijaz | Shahid |
| Sadia | Waseem | Adil |
| Faiza | Jamil | Iqbal |
| Saira | Farhan | Ashfaq |
| Komal | Nasir | Javed |
| Irfa | Bilal | Munir |
| Iqra | Nadir | Shafiq |
| Maria | Masood | Akram |
| Sania | Junaid | Ashraf |
| Irum | Hamid | Waqar |

**Figure A.1. Household income distribution in Karachi**

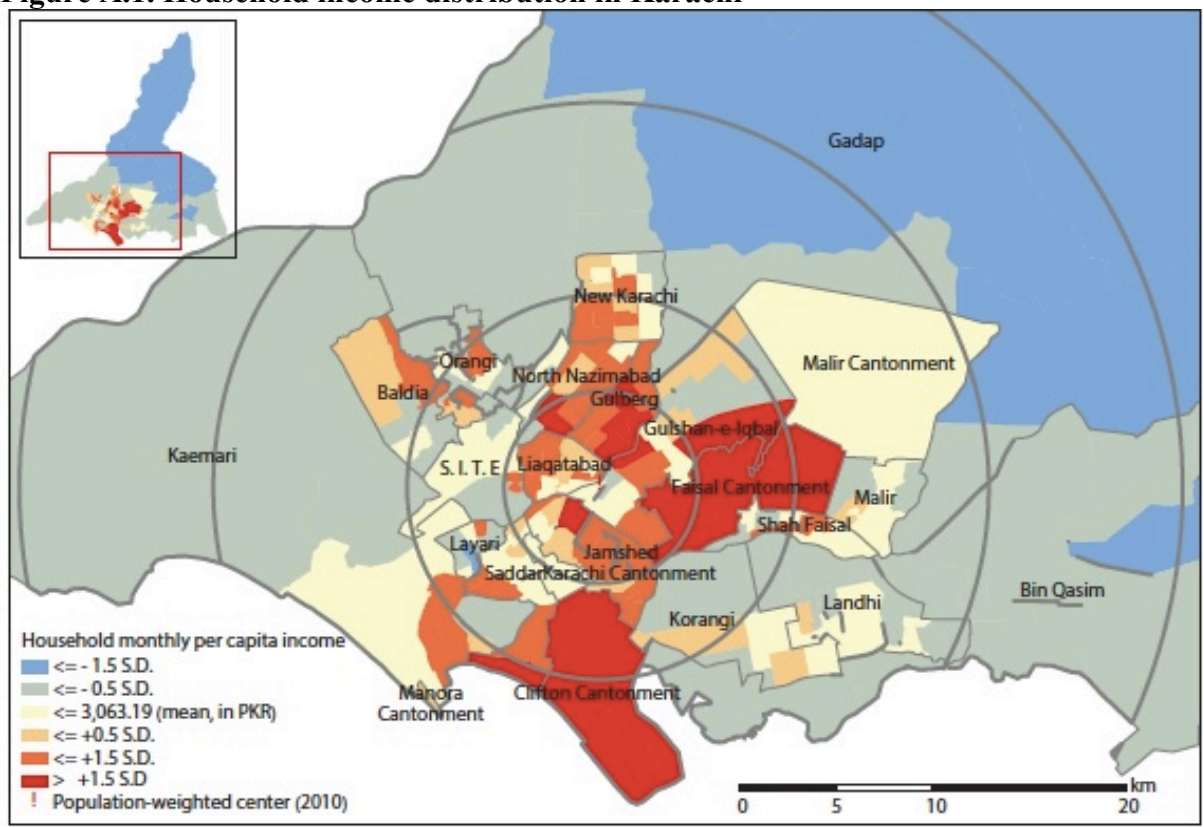

Note: Areas in red represent high-income neighborhoods, while those in grey and blue indicate low-income areas. (Source: World Bank, 2018)

**Table A.2. Socioeconomic categorization of neighborhoods by income and property values**

| Neighborhood | Average property values (PKR/Sq. ft.) | Household monthly per capita income (PKR, 2010-11) | Socioeconomic category |
|---|---|---|---|
| High-income: | | | |
| Clifton | 54,429 | 22,000 - 30,000 | High |
| Faisal Cantonment | 49,333 | 20,000 - 28,000 | High |
| Defence Housing Authority (DHA) Karachi | 33,000 | 18,000 - 25,000 | High |
| Low-income: | | | |
| Malir-Model Colony | 22,530 | 6,000 - 10,000 | Low-Mid |
| Korangi | 18,000 | 4,000 - 6,000 | Low-Mid |
| Landhi | 15,570 | 3,500 - 5,500 | Low |

Notes: Socioeconomic classification is based on household monthly per capita income and average property values. Income estimates are drawn from the Karachi Household Survey conducted by the Japan International Cooperation Agency (JICA), as cited in the World Bank's Karachi: Transforming the City into a Livable and Competitive Megacity report (2018, p. 30). Property values were extracted from Zameen.com, a leading real estate portal in Pakistan, and updated to reflect market valuations as of 2025.
Source: https://www.zameen.com/index/buy/houses/karachi-2/

**Table A.3. Mean callback rates (%) by gender**

| Group | N | Mean callback rate (%) | Std. Error (%) | 95% Confidence interval |
|-------|---|------------------------|----------------|--------------------------|
| **Men** | 1,016 | 7.87 | 0.85 | [6.22%, 9.53%] |
| **Women** | 1,016 | 6.10 | 0.75 | [4.63%, 7.57%] |
| **Comparison: Men vs. Women** | 2032 | 1.77 | 1.13 | [-0.44%, 3.99%] |

Notes: Two-sample test of proportions results: Men vs. women: z=1.566, p-value (two-tailed)=0.117
Pooled standard deviation (SD) of callback rate= 0.255
Pooled minimum detectable effect (MDE)[39]= 3.16 percentage points (Approx.)

Given a total sample size of N=2,032 (1,016 participants per gender group), a two-tailed significance level of $\alpha$=0.05, and statistical power of 80% (1−$\beta$=0.80), we can calculate the minimum detectable effect (MDE) for differences in callback rates between men and women. The observed callback rates are $p_1$=7.87% for men and $p_2$=6.10% for women, corresponding to an observed difference of 1.77 percentage points (nearly 29% higher callback rate for men relative to women).

The pooled proportion, under the null hypothesis of no difference, is estimated as p=$\frac{p_1 + p_2}{2}$= 6.98%. Using the standard normal critical values for a 5% significance level ( $\frac{z_\alpha}{2}$= 1.96) and 80% power ($z_\beta$=0.84), the MDE is calculated with the following formula:

MDE= ( $\frac{z_\alpha}{2} + z_\beta$) × $\sqrt{\frac{2p(1-p)}{n}}$

where n=1,016 is the sample size per group.[40] Substituting the values, we find MDE of 0.0316, or about 3.16 percentage points. This implies that with the current sample size, differences smaller than approximately 3.16 percentage points in callback rates cannot be detected reliably with 80% power, explaining the non-significant result for the observed 1.77 percentage point difference. Due to a lack of sufficient statistical power, we cannot reliably detect effect sizes smaller than 3.16 percentage points, such as the observed difference of 1.77 percentage points. However, the non-significant results do not prove there is no gender bias, but rather that any bias smaller than the respective MDE cannot be reliably detected with these sample sizes.

**Table A.4. Gender, neighborhood, and proximity signals: Restricted sample**

| Callback rate | Coef. | Std.Err. | t-value | p-value | [95% Conf | Interval] | Sig |
|---------------|-------|----------|---------|---------|-----------|-----------|-----|
| Gender: Man | 0.028 | 0.015 | 1.84 | 0.066 | -0.002 | 0.057 | * |
| Neighborhood: High-income | 0.034 | 0.015 | 2.20 | 0.028 | 0.004 | 0.064 | ** |
| Distance | -0.003 | 0.001 | -2.50 | 0.012 | -0.006 | -0.001 | ** |
| Constant | 0.090 | 0.021 | 4.24 | 0.000 | 0.048 | 0.132 | *** |

| | | | | |
|---|---|---|---|---|
| Mean dependent var | 0.077 | SD dependent var | 0.266 |
| R-squared | 0.013 | Number of Obs. | 1236.000 |
| F-test | 4.700 | Prob > F | 0.003 |
| Akaike crit. (AIC) | 229.802 | Bayesian crit. (BIC) | 250.280 |

*** p<0.01, ** p<0.05, * p<0.1

---

[39] MDE defines the smallest true effect the study can detect with adequate power; effects smaller than that tend to be statistically insignificant.
[40] Since MDE is derived from a two-sample comparison of proportions.

**Table A.5. Estimates of the probability of receiving a callback for a job interview (with interaction effects)**

|  | (1) | (2) | (3) |
|---|---|---|---|
| Gender (1= Man, 0= Woman) | 0.107*** | 0.028*** | 0.109** |
|  | (0.036) | (0.038) | (0.043) |
| Neighborhood (1= high-income, 0= low-income) | 0.032** | 0.052 | 0.053** |
|  | (0.015) | (0.039) | (0.025) |
| Distance | -0.0005 | -0.003** | 0.0004 |
|  | (0.002) | (0.001) | (0.002) |
| Gender × Neighborhood |  |  | -0.001 |
|  |  |  | (0.003) |
| Distance × Gender | -0.006** |  | -0.006** |
|  | (0.002) |  | (0.002) |
| Distance × Neighborhood |  | -0.0015 | -0.002 |
|  |  | (0.003) | (0.003) |
| Control for firm size | Yes | Yes | Yes |
| Control for economic sector of activity | Yes | Yes | Yes |
| Constant | 0.165*** | 0.194*** | 0.152** |
|  | (0.058) | (0.053) | (0.059) |
| Mean of Dep. Var. | 0.077 | 0.077 | 0.077 |
| Observations | 1236 | 1236 | 1236 |
| R-squared | 0.02 | 0.02 | 0.022 |

Note: Robust standard errors are in parentheses: *** $p<0.01$, ** $p<0.05$, * $p<0.1$

**Table A.6. Mean callback rates (%) by firm size**

| Callback Status | Number of employees | | | | |
|---|---|---|---|---|---|
|  | 1-10 | 11-50 | 51-100 | 100-above | Total |
| 0 (Not invited) | 358 | 768 | 127 | 246 | 1499 |
|  | 93.23 | 92.75 | 90.71 | 94.62 | 92.99 |
| 1 (Invited) | 26 | 60 | 13 | 14 | 113 |
|  | 6.77 | 7.25 | 9.29 | 5.38 | 7.01 |
| Total | 384 | 828 | 140 | 260 | 1612 |
|  | 100.00 | 100.00 | 100.00 | 100.00 | 100.00 |

Note: First row indicates frequencies; second row in parentheses indicates column percentages.

**Table A.7. Categorization of economic sectors of activity[41]**

|                                                    | Freq. | Percent | Cum.   |
|----------------------------------------------------|-------|---------|--------|
| Manufacturing                                      | 40    | 2.01    | 2.01   |
| Wholesale and retail trade                         | 60    | 3.02    | 5.03   |
| Transportation and storage                         | 40    | 2.01    | 7.04   |
| Information and communication                      | 1472  | 74.04   | 81.09  |
| Professional, scientific and technical activities  | 312   | 15.69   | 96.78  |
| Financial and other services                       | 64    | 3.22    | 100.00 |

[41] While most companies in this study primarily operate in the IT sector, some also have secondary operations in other sectors. While the number of observations in sectors outside of IT is relatively low, including these sectors allows to control for any residual sector-specific effects that might affect the main analysis.